\documentclass[review]{elsarticle}

\usepackage{lineno,hyperref}
\modulolinenumbers[5]

\journal{Journal of \LaTeX\ Templates}

\usepackage{amsmath}
\usepackage{amsfonts}
\usepackage{amssymb}

\begin{document}

\begin{frontmatter}

\title{Constraints on parity violation from ACTpol and forecasts for forthcoming CMB experiments}

\author[Ferrara,IASF]{Diego Molinari\corref{mycorrespondingauthor}}
\cortext[mycorrespondingauthor]{Corresponding author}
\ead{molinari@iasfbo.inaf.it}

\author[IASF,INFN]{Alessandro Gruppuso}

\author[Ferrara]{Paolo Natoli}

\address[Ferrara]{Dipartimento di Fisica e Scienze della Terra, Universit\`a degli Studi di Ferrara, and INFN, Sezione di Ferrara, Via Saragat 1, I-44122 Ferrara, Italy}

\address[IASF]{INAF-IASF Bologna, Istituto di Astrofisica Spaziale e Fisica Cosmica di Bologna, Istituto Nazionale di Astrofisica, via Gobetti 101, I-40129 Bologna, Italy}

\address[INFN]{INFN, Sezione di Bologna, Via Irnerio 46, I-40126 Bologna, Italy}
\begin{abstract}
We use the ACTpol published cosmic microwave background (CMB) polarization data to constrain cosmological birefringence, a tracer of parity violation beyond the standard model of particle physics. To this purpose, we employ all the polarized ACTpol spectra, including the cross-correlations between temperature anisotropy and B mode polarization (TB) and between E mode and B mode (EB), which are most sensitive to the effect. We build specific, so-called D-estimators for birefringence and assess their performances and error budgets by using realistic Monte Carlo simulations based on the experimental characteristics provided by the ACTpol collaboration. We determine the optimal multipole range for our analysis to be $250 < \ell < 3025$ over which we find a null result for the uniform birefringence angle $\alpha = 0.29^\circ \pm 0.28^\circ$ (stat.) $\pm 0.5^\circ$ (syst.), the latter uncertainty being the estimate published by the ACTpol team on their global systematic error budget. We show that this result holds consistently when other multipole ranges are considered. Finally, we forecast the capability of several forthcoming ground based, balloon and space borne CMB experiments to constrain the birefringence angle, showing, e.g., that the proposed post-Planck COrE satellite mission could in principle constrain $\alpha$ at a level of 10 arcsec, provided that all systematics are under control. Under the same circumstances, we find the COrE constraints to be at least 2 or 3 times better than what could ideally be achieved by the other experiments considered. 
\end{abstract}


\end{frontmatter}

\linenumbers

\section{Introduction}

Observations of the CMB radiation from third generation experimental efforts such as the Planck \cite{planck2014-a01} satellite and the ground based BICEP 2/Keck \cite{pb2015}, ACT \cite{ACT2013} and SPT \cite{SPT2011} observatories have ushered a new era in precision cosmology. 
Measurements of the temperature anisotropy pattern has reached very high signal-to-noise ratios and are basically limited by the sky either because of cosmic variance or because of residual errors from small scale foregrounds \cite{planck2014-a13}.
The experiments consider a wide range of angular scales depending on the instrument, i.e. $2 \leq \ell \lesssim 2500$ for Planck \cite{planck2014-a13}, $300 \lesssim \ell \lesssim 8500$ for ACT \cite{AdvACTpol2015}, $2000 \lesssim \ell \lesssim 13000$ for SPT \cite{SPTpol2014}.
They have also released the first CMB polarization measurements with significantly high signal-to-noise ratios to effectively exploit cosmological information in the polarized components of the CMB.
New observational campaigns from ground based experiments (BICEP 3 \cite{Bicep32014}, Advanced ACT \cite{ACTpol2014}, SPTpol \cite{SPTpol2015}), balloon-borne experiments (LSPE \cite{LSPE2012}) and satellite proposals (LiteBIRD \cite{Litebird2014}, COrE \cite{Core+2014}) are moving towards cosmic variance limited measurements also in polarization. 
This effort is motivated, other than by the will to better constrain the cosmological standard model, also by the unique possibility offered from CMB polarization to test for new physics.

Cosmic birefringence, the in vacuo rotation of the photons polarization direction during propagation \cite{Carroll1990}, is already considered as a standard tracer of parity violating mechanisms beyond the Maxwell Lagrangian. Several theoretical models exhibit such an effect including Chern-Simons type interactions in the electromagnetic Lagrangian \cite{Carroll1990}, a quintessence field \cite{Carroll1998,Giovannini2005}, axion-like particles coupled with the electromagnetic field \cite{Finelli2009}, spatial anisotropies during evolution of perturbations \cite{Bartolo2015}.
No detection of in vacuo birefringence has been claimed today: the effect has been constrained by laboratory experiments to be small \cite{PVLAS2016}. Astrophysical probes are very good candidates for precision measurements because of the long journey engaged by cosmological photons. Distant radio galaxies are  widely studied to test for parity violations (see \cite{Carroll1998,Alighieri2010,Kamionkowski2010} and references therein).

Cosmic birefringence may also affect CMB photons, causing a non-zero cross-correlation between temperature anisotropies and curl-like polarization patterns or B modes, and between gradient-like polarization patterns, or E modes, and B modes. These correlations can be parametrized by an angle $\alpha$ known as the birefringence angle\footnote{We use the customary convention used by the CMB community for the Q and U Stokes parameters, see e.g. \\ http://wiki.cosmos.esa.int/planckpla2015/index.php/Sky\_temperature\_maps}. In this paper we restrict to the case of a uniform rotation angle $\alpha$, whose effect on the CMB spectra is \cite{Lue1999, Li2009}:
\begin{eqnarray}
C_{\ell}^{TE,obs} &=& C_{\ell}^{TE} cos(2\alpha) \\
\label{rotation1}
C_{\ell}^{TB,obs} &=& C_{\ell}^{TE} sin(2\alpha) \\
\label{rotation2}
C_{\ell}^{EE,obs} &=& C_{\ell}^{EE} cos^2(2\alpha) + C_{\ell}^{BB} sin^2(2\alpha) \\
\label{rotation3}
C_{\ell}^{BB,obs} &=& C_{\ell}^{BB} cos^2(2\alpha) + C_{\ell}^{EE} sin^2(2\alpha) \\
\label{rotation4}
C_{\ell}^{EB,obs} &=& \frac{1}{2} (C_{\ell}^{EE}-C_{\ell}^{BB}) sin(4\alpha)
\label{rotation5}
\end{eqnarray}
where $C_{\ell}^{XY,obs}$ and $C_{\ell}^{XY}$ are the observed and the primordial (i.e. unrotated) angular power spectra (APS) of the $XY$ spectrum ($X$,$Y = T$, $E$ or $B$) respectively. The most constraining CMB spectra are TB and EB that are predicted to be zero by the standard cosmological model \cite{Lue1999}.
Existing limits on $\alpha$  are all compatible with a null effect, for a review see \cite{Alghieri2015,Kaufman2016}. Recent analyses of uniform rotation angle obtained with Planck data can be found in \cite{Contaldi2015,Gruppuso2015}. For anisotropic birefringence angle constraints see \cite{Gluscevic2012, Polarbear2015}. Studies of the interplay between primordial B modes and birefringence angles are e.g. given in \cite{Zhaobis2014,Zhao2014}.

In this paper we provide constraints of the birefringence angle employing an independent analysis of the polarization data recently published by the ACTpol Collaboration \cite{ACTpol2014}. Moreover we give forecasts of $\alpha$ for future CMB experiments.
The structure of this paper is the following: in Section \ref{Metod} we present the method employed to estimate the birefringence angle; in Section \ref{DataandSimulations} we describe the details of the data analysis and the procedure followed to generate realistic Monte Carlo (MC) simulations. We present our results from the ACTpol analyses in Section \ref{Results} whilst in Section \ref{Forecasts} we provide our forecasts for several forthcoming and proposed CMB experiments. We draw our main conclusions in Section \ref{Conclusions}.

\section{Birefringence estimators}\label{Metod}

To find an estimate, $\tilde{\alpha}$, for the birefringence angle, $\alpha$, we consider the so-called D-estimators (see \cite{Gluscevic2009,Yadav2009,Wu2009,Gruppuso2012,Zhao2015,Gruppuso2016} for details) defined as:
\begin{eqnarray}
D_{\ell}^{TB}(\tilde{\alpha}) &=& C_{\ell}^{TB,obs}cos(2\tilde{\alpha}) - C_{\ell}^{TE,obs}sin(2\tilde{\alpha}) \, ,
\label{estimators1}\\
D_{\ell}^{EB}(\tilde{\alpha}) &=& C_{\ell}^{EB,obs}cos(4\tilde{\alpha}) - \frac{1}{2}(C_{\ell}^{EE,obs}-C_{\ell}^{BB,obs})sin(4\tilde{\alpha}) \, .
\label{estimators2}
\end{eqnarray}
When $\tilde{\alpha}=\alpha$ the expectation value of Eq.~(\ref{estimators1}) and Eq.~(\ref{estimators2}) is zero. This can be easily seen by replacing Eq.~(\ref{rotation1})-(\ref{rotation5}) in Eq.~(\ref{estimators1}),(\ref{estimators2}) \cite{Gruppuso2016}. 

In order to find $\tilde{\alpha}$, we minimize the following figure of merit:
\begin{equation}\label{chisq}
\chi_{X}^2(\tilde{\alpha}) = \sum_{\ell\ell'} D_{\ell}^{X}M_{\ell\ell'}^{XX^{-1}}D_{\ell'}^{X}
\end{equation}
where $X$ stands for either TB or EB and the matrix $M_{\ell\ell'}^{XX}=\langle D_{\ell}^{X}D_{\ell'}^{X}\rangle$ is the covariance matrix of the considered D-estimator.
Instead of using Eq. (\ref{chisq}) separately for either of the D-estimators, one may consider a joint figure of merit, $D_{TBEB}$, for the vector $(D_{\ell}^{TB},D_{\ell}^{EB})$ that should take into account also the correlations between TB and EB spectra \cite{Gruppuso2016}.

The analysis can be also performed after having divided the entire multipole range in small intervals, and then by minimizing the figure of merit for each interval. This probes the dependency of $\tilde{\alpha}$ from the angular scale:
\begin{equation}\label{chisq2}
\chi_{X,\ell}^2(\tilde{\alpha})=\sum_{\ell'} D_{\ell}^{X}M_{\ell\ell'}^{XX^{-1}}D_{\ell'}^{X} \, .
\end{equation}
Resumming over $\ell$ we obtain again Eq. (\ref{chisq}).

\section{Data and simulations}\label{DataandSimulations}

\begin{figure}[h]
\centering
\includegraphics[scale=0.8]{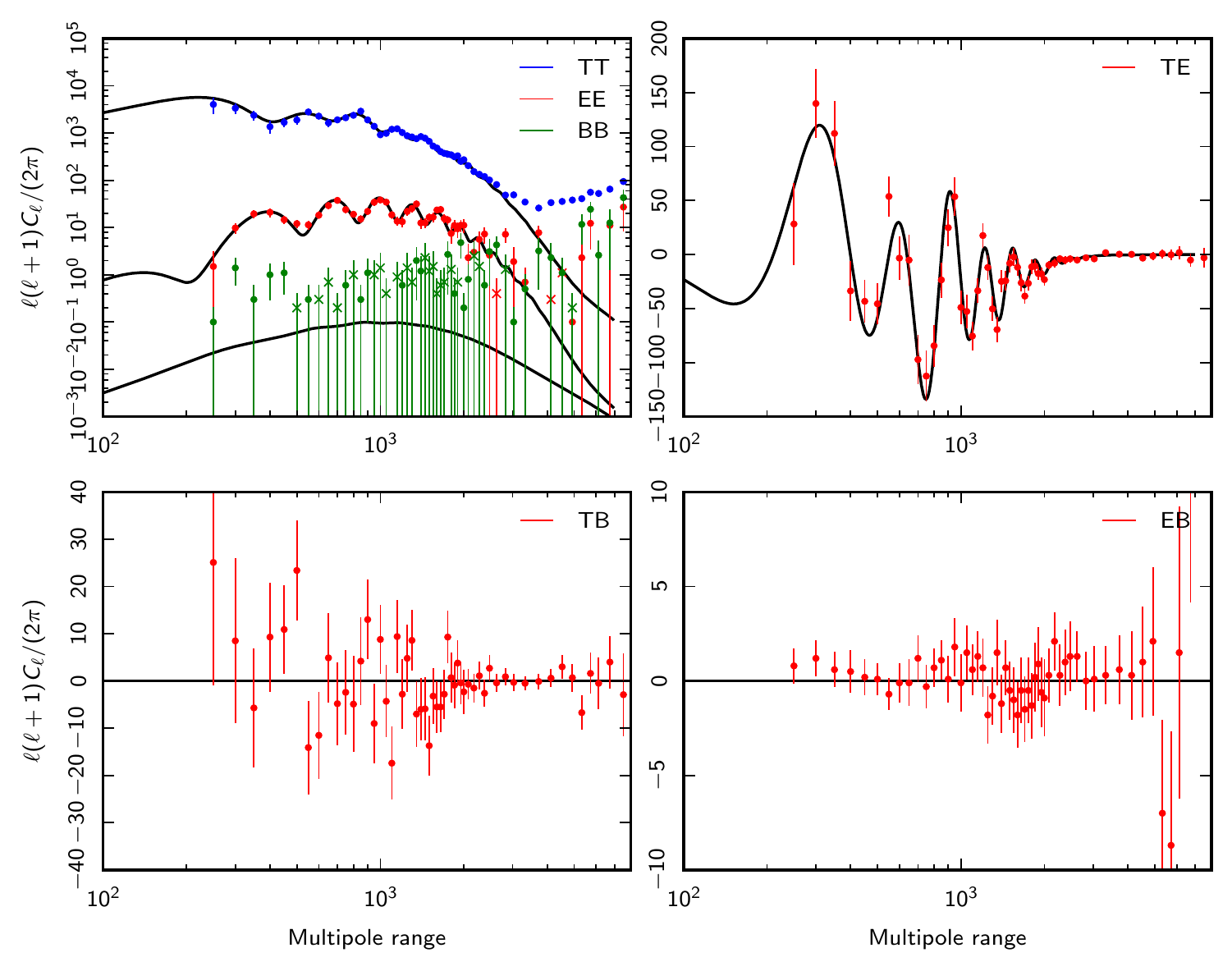}
\caption{\footnotesize{ACTpol 2014 spectra described in \cite{ACTpol2014}. Spectra are as labelled. The black curve is the Planck best fit $\Lambda$CDM model. Notice the marked deviation from model in the TT APS for $\ell \gtrsim 3000$. This is due to contamination from small scales foregrounds. It has no effect on our analysis since we do not consider TT. A possible bias in EE at small scales has been constrained in \cite{ACTpol2014} to be below $3 \sigma$, putting a limit on the amplitude of polarized sources. We ignore this effect hereafter. The 'x' symbols for BB and EE flag points that happen to be negative and for which the absolute value is displayed.}}\label{ACTdata}
\end{figure}

We analyse the recent data coming from the first three months of ACTpol observations covering 270 square degrees at 146 GHz with 1.3 arcminute resolution FWHM. These data are described in \cite{ACTpol2014} and the 6 CMB APS are publicly available\footnote{Available at http://lambda.gsfc.nasa.gov/product/act/actpol\_prod\_table.cfm . Note that the ACTpol published power spectra bands have not been subject to a nulling technique to ensure that TB and EB are compatible with zero. The technique is only used to assess the systematic error in polarization orientation \cite{PrivateComm}. }. 
They cover the multipole range $225 \leq \ell \leq 8725$ in bins varying from $\Delta\ell = 50$ for $225 \leq \ell \leq 2025$ to $\Delta\ell = 800$ for $6325 \leq \ell \leq 8725$.
In Fig. \ref{ACTdata} we show the ACTpol measurements compared to the Planck best fit $\Lambda$CDM model \cite{planck2014-a15}.

In order to propagate the errors through our analysis, we generate a realistic MC of 10000 simulated CMB APS based on the Planck 2015 $\Lambda$CDM best fit model. We work directly in harmonic space; to generate a set of 6 correlated CMB APS we build, for each bin, $\hat{\ell}$, a $6$x$6$ covariance matrix, $A_{\hat{\ell}}$: 
\begin{equation}\label{covariance}
A_{\hat{\ell}} = \begin{pmatrix}
(\sigma_{\hat{\ell}}^{TT})^2 & \Delta_{\hat{\ell}}^{TTEE} & 0 & \Delta_{\hat{\ell}}^{TTTE} & 0 & 0 \\
\Delta_{\hat{\ell}}^{TTEE} & (\sigma_{\hat{\ell}}^{EE})^2 & 0 & \Delta_{\hat{\ell}}^{EETE} & 0 & 0 \\
0 & 0 & (\sigma_{\hat{\ell}}^{BB})^2 & 0 & 0 & 0 \\
\Delta_{\hat{\ell}}^{TTTE} & \Delta_{\hat{\ell}}^{EETE} & 0 & (\sigma_{\hat{\ell}}^{TE})^2 & 0 & 0 \\
0 & 0 & 0 & 0 & (\sigma_{\hat{\ell}}^{TB})^2 & \Delta_{\hat{\ell}}^{TBEB} \\
0 & 0 & 0 & 0 & \Delta_{\hat{\ell}}^{TBEB} & (\sigma_{\hat{\ell}}^{EB})^2 \\
\end{pmatrix}
\end{equation}
where we use as diagonal elements the published ACTpol variances, $(\sigma_{\hat{\ell}}^{X})^2$, $(X=TT, EE, BB, TE, TB$ or $EB)$. 
Since we consider also primordial BB spectrum, in addition to the correlations between the spectra TT and EE ($\Delta_{\hat{\ell}}^{TTEE}$), TT and TE, ($\Delta_{\hat{\ell}}^{TTTE}$), and EE and TE, ($\Delta_{\hat{\ell}}^{EETE}$) as considered in \cite{Zaldarriaga1997,Kamionkowski1997}, we take into account also correlations between TB and EB, ($\Delta_{\hat{\ell}}^{TBEB}$). 
Assuming the diagonal elements as given, the non-zero off diagonal terms of the $A_{\hat{\ell}}$ matrix can be written as:
\begin{eqnarray}\label{Knox}
\Delta_{\hat{\ell}}^{TTEE} &=& \frac{2}{(2\hat{\ell}+1)f_{sky}}(C_{\hat{\ell}}^{TE})^2 \\
\Delta_{\hat{\ell}}^{TTTE} &=& \sqrt{\frac{2}{(2\hat{\ell}+1)f_{sky}}}(C_{\hat{\ell}}^{TE}\sigma_{\hat{\ell}}^{TT}) \\
\Delta_{\hat{\ell}}^{EETE} &=& \sqrt{\frac{2}{(2\hat{\ell}+1)f_{sky}}}(C_{\hat{\ell}}^{TE}\sigma_{\hat{\ell}}^{EE}) \\
\Delta_{\hat{\ell}}^{TBEB} &=& \sqrt{\frac{1}{2(2\hat{\ell}+1)f_{sky}}}(C_{\hat{\ell}}^{TE}\sigma_{\hat{\ell}}^{BB})
\end{eqnarray}
where $C_{\hat{\ell}}^{TE}$ is taken from the Planck 2015 $\Lambda$CDM best fit model and $f_{sky}$ is the effective sky fraction used for the APS estimation. While this value is not quoted by the ACTpol team, it can be inferred from the magnitude of the published error bars. In any case, we find that the impact of the off-diagonal elements of the covariance matrix is weak. We also assume that the noise does not correlate among temperature and polarization. Note that the above procedure allows one to account not only for white noise and cosmic variance, but also for other (systematic) effects that are already present in the published error bars of the spectra of ACTpol data \cite{ACTpol2014}.

To build a MC set of simulated spectra, we use the Cholesky factor of Equation \ref{covariance} assuming a reference $\Lambda$CDM model. We then compare the estimates $\tilde{\alpha}$ from the MC simulations with those from real data.

\begin{figure}
\centering
\includegraphics[scale=0.8]{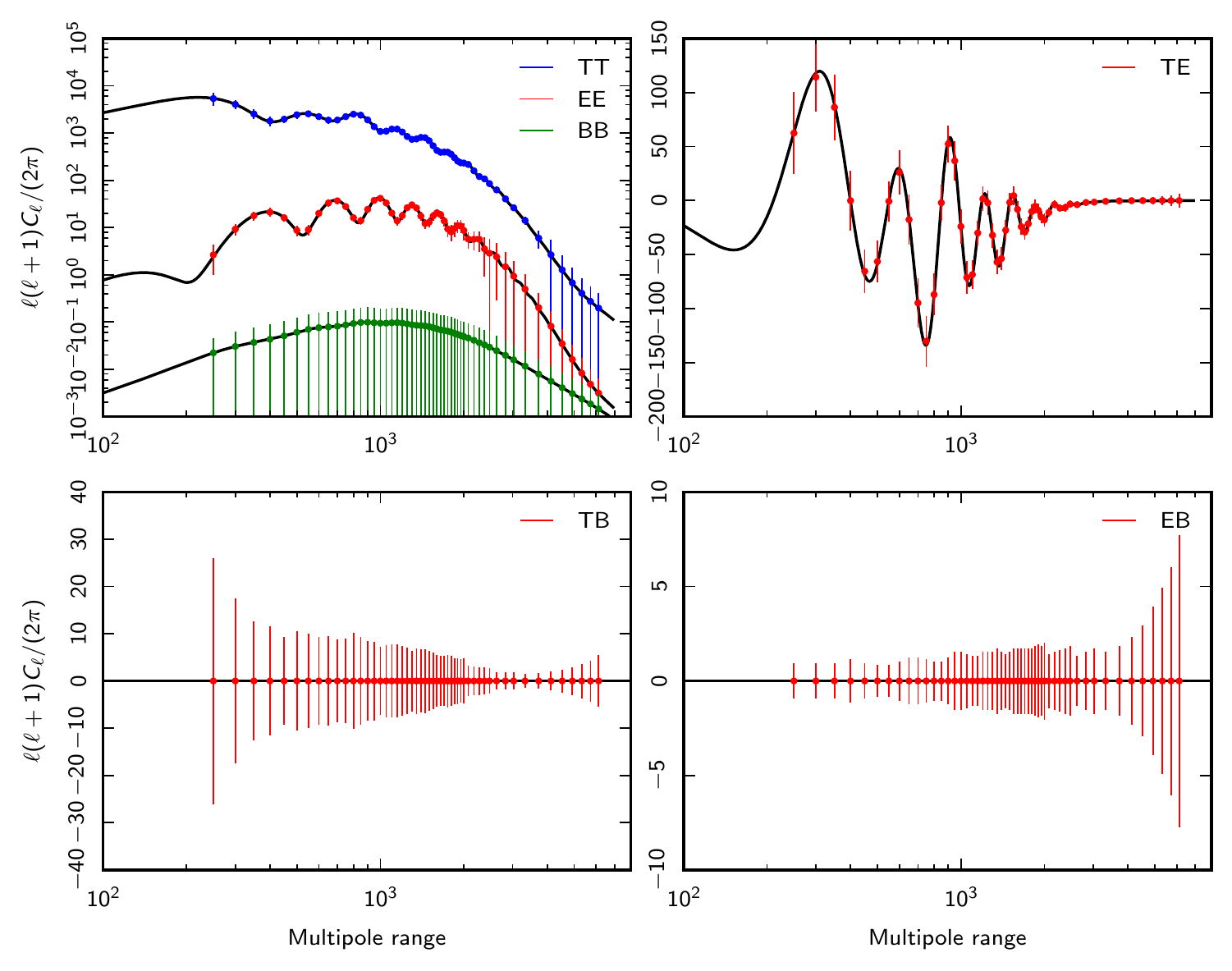}
\caption{\footnotesize{Average and dispersion of the 10000 simulated APS. Clockwisely the TT, EE and BB APS ({\it top left panel}), the TE APS ({\it top right panel}), the EB APS ({\it bottom right panel}) and the TB APS ({\it bottom left} panel).}}\label{MCdata}
\end{figure}

In Fig. \ref{MCdata} we show the mean and standard deviation of the 10000 MC simulated APS.

\section{Results}\label{Results}

We compute and minimize the figure of merit of Eq. (\ref{chisq}) for the ACTpol data and the MC simulations, considering the multipole range $[250,3025]$. The minimum multipole is dictated by the ACTpol data, while the choice of the maximum multipole is derived by considerations of the signal-to-noise ratio, highlighted in Appendix A.

\begin{figure}
\centering
\includegraphics[scale=0.6]{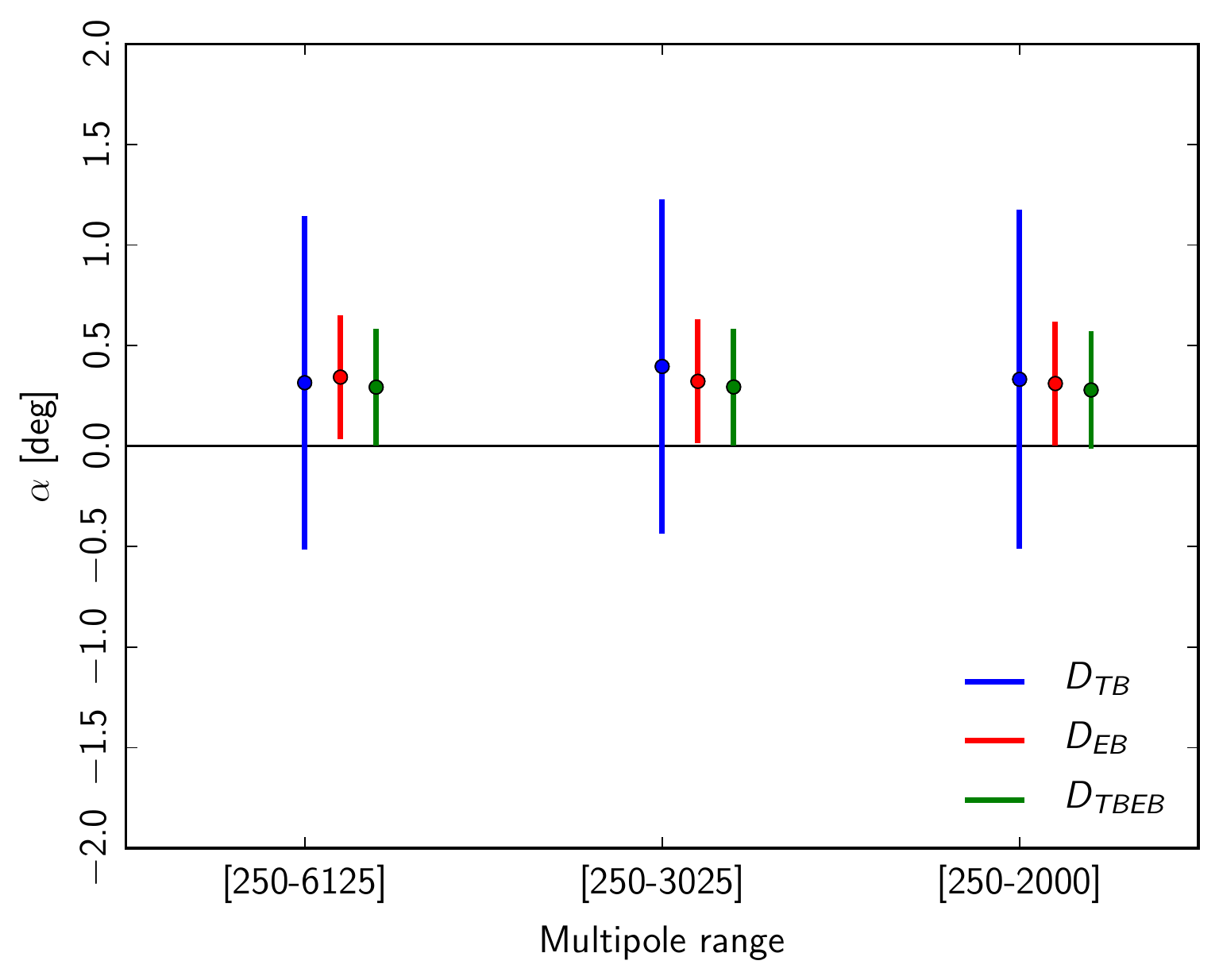}
\caption{\footnotesize{Estimates of the birefringence angle from $D_{TB}$ (blue bars), $D_{EB}$ (red bars) or $D_{TBEB}$ (green bars) for several multipole ranges. }}\label{alpha_allrange}
\end{figure}

In Fig. \ref{histograms} we show the empirical distribution of the estimated birefringence angle for $D_{TB}$ (left panel), $D_{EB}$ (central panel) and $D_{TBEB}$ (right panel) of the MC. The dark blue bar highlights the value derived from the ACTpol data. In Fig. \ref{alpha_allrange} we show the results obtained using the $D_{TB}$ (blue bars), $D_{EB}$ (red bars) or $D_{TBEB}$ (green bars) estimators, considering several multipole ranges.

\begin{figure}
\centering
\includegraphics[scale=0.8]{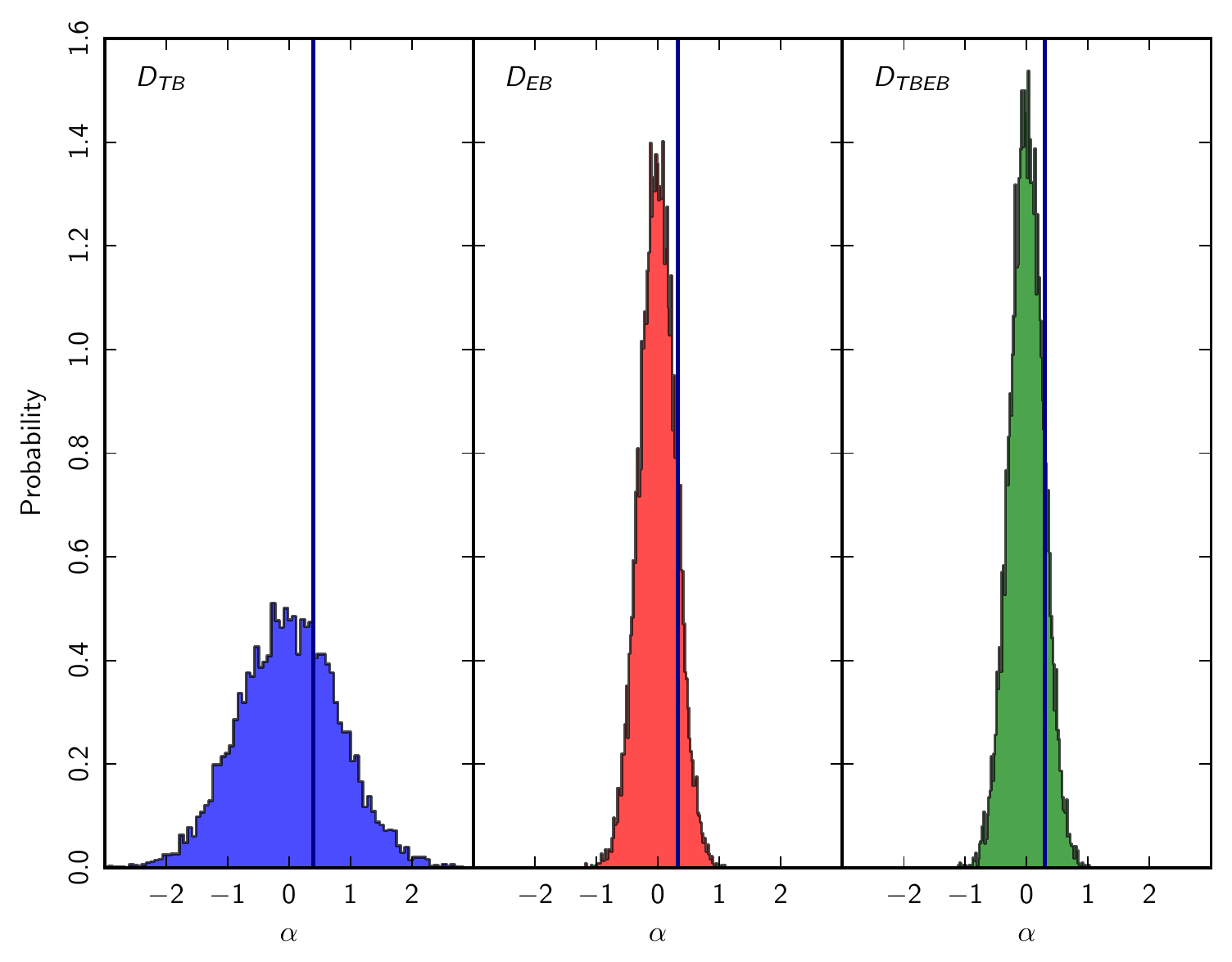}
\caption{\footnotesize{Normalized distributions of birefringence angles estimated from the MC simulations using $D_{TB}$ (left panel), $D_{EB}$ (central panel) and $D_{TBEB}$ (right panel) compared with the ones derived from the ACTpol data (dark blue). }}\label{histograms}
\end{figure}

\begin{table}
\centering
\begin{tabular}{|c|c|c|c|}
\hline
Multipole range & $D_{TB}$ [deg] & $D_{EB}$ [deg] & $D_{TBEB}$ [deg] \\ 
\hline
$250-6125$ & $0.32 \pm 0.82$ & $0.34 \pm 0.29$ & $0.29 \pm 0.28$ \\
\hline
$250-3025$ & $0.40 \pm 0.82$ & $0.32 \pm 0.29$ & $0.29 \pm 0.28$ \\
\hline
$250-2000$ & $0.33 \pm 0.83$ & $0.31 \pm 0.30$ & $0.28 \pm 0.28$ \\
\hline
\end{tabular}
\caption{Estimated birefringence angle using $D_{TB}$, $D_{EB}$ or $D_{TBEB}$ for several multipole ranges (first column). Note the overall consistency of the estimated angles.}\label{table}
\end{table}

In addition to the range $[250,3025]$ we also take into account the multipole interval $[250,6125]$ which comprises the entire set of useful power spectra bands provided by the ACTpol. Moreover we consider the multipole range $[250,2000]$ for consistency test. All our constraints are well compatible each other as reported in Table \ref{table}. We also addressed the impact of the cross-correlations between TB and EB spectra in the $D_{TBEB}$ estimator finding it negligible.

Our best, representative estimate is $\tilde{\alpha} = (0.29 \pm 0.28)^\circ$ (all the statistical errors given in this paper are at $1\sigma$ level). It is in good agreement with previous analyses \cite{Mei2015,Zhao2015} showing an estimated birefringence angle consistent with zero at about $1\sigma$. The error quoted above is purely statistical and does not include a systematic contribution estimated in \cite{ACTpol2014} at $0.5^\circ$. This error is of the same order of magnitude as the statistical error derived above so both have to be considered together when quoting the ACTpol constraints on $\alpha$.

\subsection{Consistency tests across multipole ranges}

\begin{figure}
\centering
\includegraphics[scale=0.6]{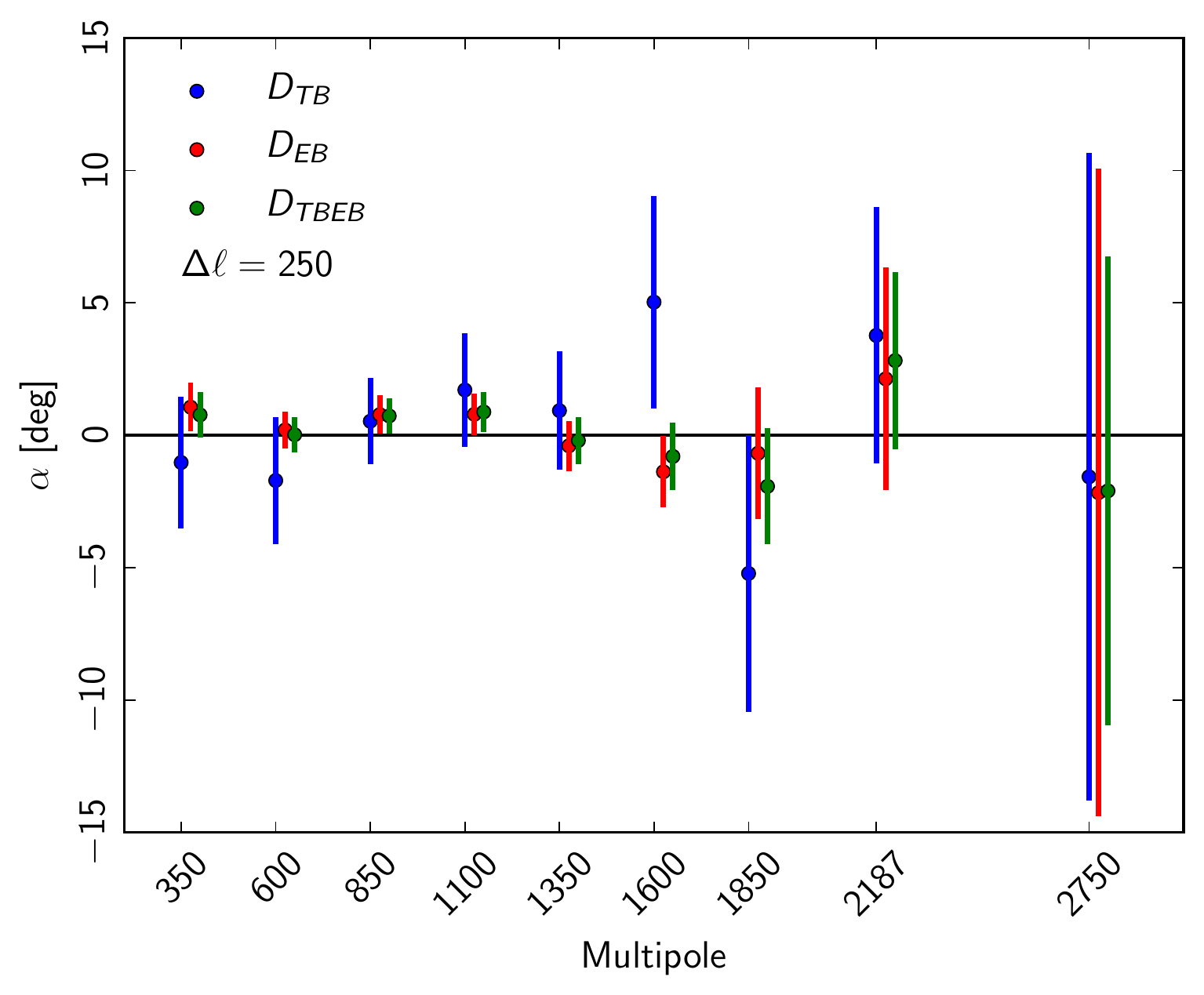}
\caption{\footnotesize{Extraction of the uniform birefringence angle considering the multipole range $[250,3025]$ divided in 9 subintervals with $\Delta \ell = 250$ up to $\ell = 1950$, $\Delta \ell = 400$ up to $\ell = 2375$ and $\Delta \ell = 600$ up to $\ell = 3025$. }}\label{alpha_spettro}
\end{figure}

The D-estimators can be used to extract $\tilde{\alpha}$ in different multipole ranges as in Eq. (\ref{chisq2}). In Fig. \ref{alpha_spettro} we show estimates of the uniform rotation angle $\tilde{\alpha}$ as obtained at different angular scales by dividing the multipole range $[250,3025]$ in 9 intervals, i.e. in bins of $\Delta\ell = 250$ up to $\ell = 1950$, $\Delta \ell = 400$ up to $\ell = 2375$ and $\Delta \ell = 600$ up to $\ell = 3025$. Errors blow up significantly beyond this value.
We find that all the estimates of the uniform birefringence angle do not significantly deviate from zero in any subinterval considered. This shows that the assumption of a uniform rotation angle continues to hold also when subsets of the full multipole range are considered.

\section{Forecasts for future experiments}\label{Forecasts}

\begin{table}
\centering
\begin{tabular}{|c|c|c|c|c|c|}
\hline
Description & $\nu$ & $f_{sky}$ & Beam & $\sigma_T$ & $\sigma_P$ \\
\hline
 & [GHz] & [\%] & [arcmin] & [$\mu K$-arcmin] & [$\mu K$-arcmin] \\
\hline
Advanced ACT\cite{AdvACTpol2015} & 150 & 50 & 1.4 & 7.0 & 9.9\\
\hline
LSPE\cite{LSPE2012} & 95 & 20 & 110 & 7.8 & 11.0 \\
\hline
LiteBIRD\cite{Litebird2014} & 140 & 80 & 32 & 3.7 & 5.2 \\
\hline 
COrE-like sat.\cite{Core+2014} & 135 & 80 & 7.8 & 2.63 & 4.55 \\
\hline
Ideal (r=0.1) & - & 100 & - & - & - \\
\hline
Ideal (r=0.01) & - & 100 & - & - & - \\
\hline
\end{tabular}
\caption{\footnotesize{List of the forecasts considered. We consider only the frequency with most favourable sensitivity to CMB polarization. For the definition of temperature and polarization sensitivities see subsection \ref{methodology}. }}\label{tab:forecasts}
\end{table}

In this Section we derive forecasts for forthcoming space-born and sub-orbital experiments using the machinery described above.

\subsection{Methodology}\label{methodology}

We build the covariance matrix by considering the explicit covariances for an idealized white noise only experiment, \cite{Zaldarriaga1997,Kamionkowski1997}, with gaussian circular beam. We allow for the possibility of having non zero primordial B modes, but assume that the primordial TB and EB spectra are zero, i.e. there is no parity violating signal from the early universe. The diagonal elements of the covariance matrix read:
\begin{align}
(\Delta C_{\ell}^{TT})^2 &= \frac{2}{(2\ell+1)f_{sky}}(C_{\ell}^{TT} + w_T^{-1}W_{\ell}^{-2})^2 \\
(\Delta C_{\ell}^{EE})^2 &= \frac{2}{(2\ell+1)f_{sky}}(C_{\ell}^{EE} + w_P^{-1}W_{\ell}^{-2})^2 \\
(\Delta C_{\ell}^{BB})^2 &= \frac{2}{(2\ell+1)f_{sky}}(C_{\ell}^{BB} + w_P^{-1}W_{\ell}^{-2})^2 \\
(\Delta C_{\ell}^{TE})^2 &= \frac{1}{(2\ell+1)f_{sky}}[(C_{\ell}^{TT} + w_T^{-1}W_{\ell}^{-2})(C_{\ell}^{EE} + w_P^{-1}W_{\ell}^{-2}) + (C_{\ell}^{TE})^2] \\
(\Delta C_{\ell}^{TB})^2 &= \frac{1}{(2\ell+1)f_{sky}}[(C_{\ell}^{TT} + w_T^{-1}W_{\ell}^{-2})(C_{\ell}^{BB} + w_P^{-1}W_{\ell}^{-2})] \\
(\Delta C_{\ell}^{EB})^2 &= \frac{1}{(2\ell+1)f_{sky}}[(C_{\ell}^{EE} + w_T^{-1}W_{\ell}^{-2})(C_{\ell}^{BB} + w_P^{-1}W_{\ell}^{-2})]
\end{align}
where $f_{sky}$ is the useful sky fraction available, $W_{\ell}$ is a (Gaussian) window function, $w_{T,P}^{-1}=\sigma_{T,P}^2$ are the inverse statistical weights per unit solid angle and $\sigma_{T,P}$ are the noise sensitivities per unit solid angle of the instrument in temperature and polarization. The non-zero off diagonal terms are:
\begin{eqnarray}
Cov(C_{\ell}^{TT},C_{\ell}^{EE}) &=& \frac{2}{(2\ell+1)f_{sky}}(C_{\ell}^{TE})^2 \\
Cov(C_{\ell}^{TT},C_{\ell}^{TE}) &=& \frac{2}{(2\ell+1)f_{sky}}[C_{\ell}^{TE}(C_{\ell}^{TT}+w_T^{-1}W_{\ell}^{-2})] \\
Cov(C_{\ell}^{EE},C_{\ell}^{TE}) &=& \frac{2}{(2\ell+1)f_{sky}}[C_{\ell}^{TE}(C_{\ell}^{EE}+w_P^{-1}W_{\ell}^{-2})] \\
Cov(C_{\ell}^{TB},C_{\ell}^{EB}) &=& \frac{1}{(2\ell+1)f_{sky}}[C_{\ell}^{TE}(C_{\ell}^{BB}+w_P^{-1}W_{\ell}^{-2})]
\end{eqnarray}

\subsection{Experimental set-up}

In Table \ref{tab:forecasts} we show the experiments we consider and their main features. All of them are multi-frequency instruments. As reference, we consider the single frequency with the most favourable sensitivity to CMB polarization. 
The new observational campaign of the ACT telescope, called Advanced ACT \cite{AdvACTpol2015}, will observe about $50\%$ of the sky for 3 years. At 150 GHz it will feature a beam of $1.4$ arcmin reaching a noise level of $7\, \mu K$-arcmin in temperature and $9.9\, \mu K$-arcmin in polarization. 
The balloon experiment LSPE \cite{LSPE2012} will observe $20\%$ of the sky between 30 GHz and 245 GHz with a telescope beam of $110$ arcmin at 95 GHz reaching a map noise level of $7.8\, \mu K$-arcmin in temperature and $11.0\, \mu K$-arcmin in polarization.
The LiteBIRD satellite\cite{Litebird2014} will observe the CMB sky between 60 GHz and 280 GHz with a beam of $32$ arcmin at 145 GHz and a sensitivity of $3.7\, \mu K$-arcmin in temperature and $5.2\, \mu K$-arcmin in polarization. 
A COrE-like satellite\cite{Core+2014} may observe the CMB sky between 45 GHz ad 795 GHz with a resolution of $7.8$ arcmin at 135 GHz and a noise level of $2.63\, \mu K$-arcmin in temperature and $4.55\, \mu K$-arcmin in polarization. 
For these last two cases we consider $f_{sky}=80\%$, a reasonable value allowing for some expected developments in the component separation techniques. 
Since the sky fraction available for CMB analysis is expected to be large for these future experiments, we consider a multipole range that differs in extension and binning size from the one employed above for ACTpol. We thus consider $\Delta\ell = 10$ up to $\ell = 500$, $\Delta\ell = 50$ up to $\ell = 2000$, $\Delta\ell = 100$ up to $\ell = 2500$ and $\Delta\ell = 200$ up to $\ell = 3000$. For the LiteBIRD and CORE we consider $\ell_{min} = 2$ since they will analyse a larger portion of the sky, for Advanced ACT and LSPE we use $\ell_{min} = 10$. We consider the signal-to-noise ratio to determine for each experiment the multipole range useful for the birefringence analysis. Details are described in Appendix A. The results show that the maximum multipole useful is $\ell_{max} = 3000$ for Advanced ACT, $\ell_{max} = 200$ for LSPE, $\ell_{max} = 600$ for LiteBIRD and $\ell_{max} = 2000$ for COrE.
Finally we consider also a full sky analysis without noise in order to quantify the limit to the sensitivity of the D-estimators from cosmic variance. For these cases, the unlensed spectra derived from the $\Lambda$CDM and no noise contribution would result in a null BB spectrum and thus null EB and TB spectra. Their errors due to cosmic variance would be therefore null as well. In order to avoid this degenerate scenario, we consider two $\Lambda$CDM extended models with two different values of the tensor-to-scalar ratio, $r$, of 0.1 and 0.01. The binning and the multipole range is taken the same as COrE and we consider $\ell_{max}=3000$.

One relevant aspect connected with forecasting is the impact of CMB weak lensing on the birefringence angle estimation. 
Weak lensing is a parity conserving effect \cite{Gruppuso2016} and therefore one would not expect it to affect the estimate and the error budget of $\alpha$. This can be proven formally by explicitly computing the $\chi^2$ of the lensed spectra \cite{Gruppuso2016}. As a consequence one can compute forecasts relying on the unlensed spectra.
However, when dealing with real data, the observed spectra will be lensed and a knowledge of the lensing kernel is required to deconvolve the effect. In practice, there will be some degree of uncertainty connected with the reconstruction of the lensing potential, which may impact the error budget on $\alpha$ or even bias its estimate.
At the sensitivity of the ACTpol data treated in this paper, one does not expect any significant impact on B-modes by weak lensing and the effect can be safely ignored \cite{ACTpol2014}.
For the forecasts considered below, which refer to high sensitivity experiments where weak lensing is indeed relevant, we implicitly assume that the lensing kernel can be perfectly deconvolved. Thus, for the following analysis, we consider MC of unlensed CMB APS. The impact on $\alpha$ of a misestimation of the lensing kernel is addressed in \cite{Gruppuso2016}.

\subsection{Results}

\begin{figure}
\centering
\includegraphics[scale=0.8]{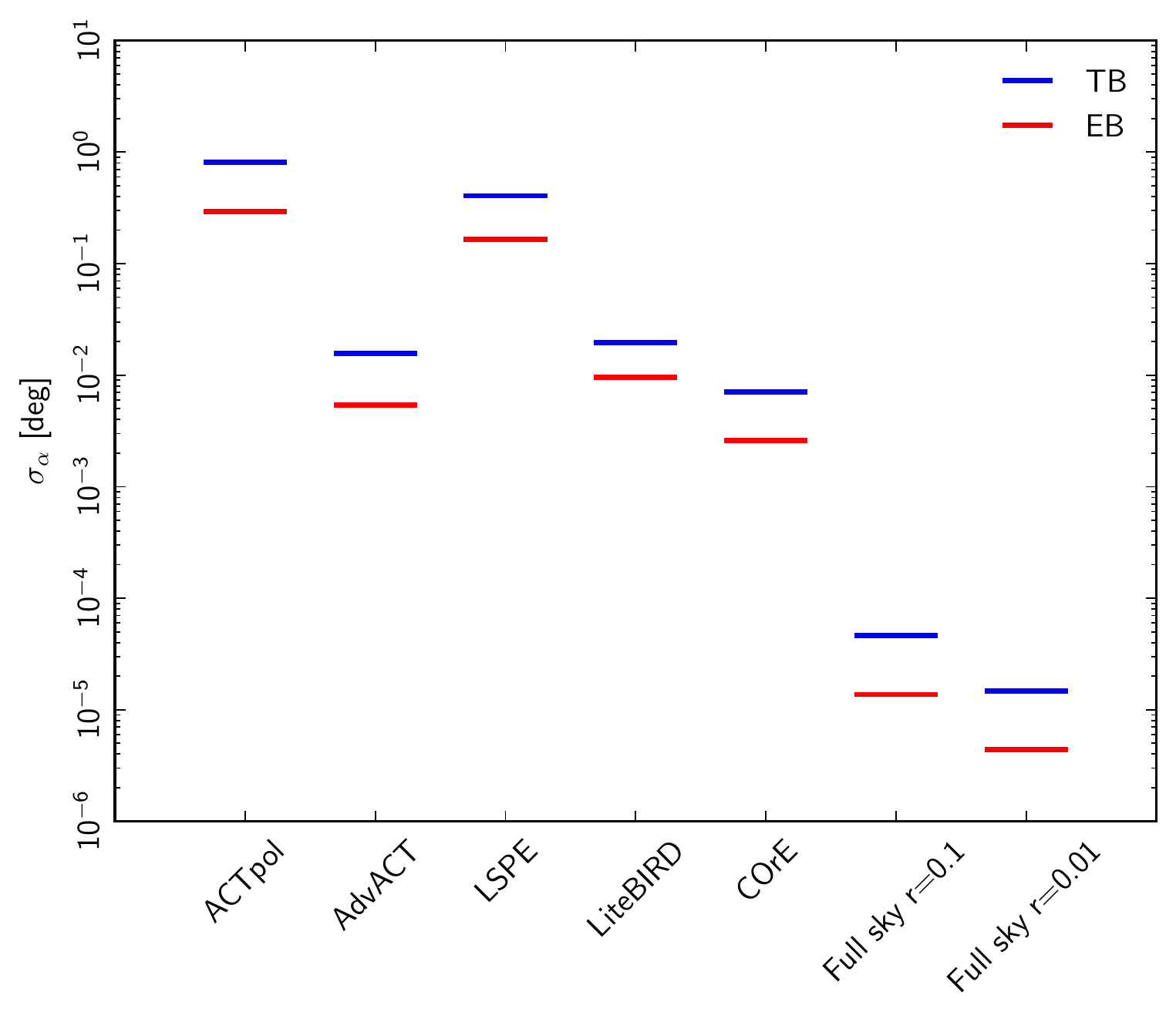}
\caption{\footnotesize{Forecasts for $1 \sigma$ errors in the measurements of the birefringence angle using the D-estimators, $D_{TB}$ (blue) and $D_{EB}$ (red) for the future experiments listed. We also quote the error estimates achievable with public ACTpol data discussed above in this paper and two ideal, cosmic variance limited, experiments assuming a value of the tensor-to-scalar ratio $r = 0.1$ or $0.01$ respectively. We do not show the $D_{EBTB}$ results because they are in practice undistinguishable from the $D_{EB}$ ones.}}\label{forecasts}
\end{figure}

We compute the figure of merit of Eq. (\ref{chisq}) using a MC of 10000 unlensed CMB APS. We report in Fig. \ref{forecasts} the expected $1 \sigma$ error on the estimate of the birefringence angle $\tilde{\alpha}$ for each experiment considered in Table \ref{tab:forecasts}. The ideal full sky cases can be interpreted as the cosmic variance limit of our D-estimators assuming two different fiducial values for $r$.

The best prospects for constraining $\alpha$ are provided by a COrE-like satellite mission which will be able to achieve a sensitivity for $\alpha$ of the level of 10 arcsec. This of course assume negligible impact of systematic effects from both the sky and the instrument. 
Under the same condition the LiteBIRD mission has a sensitivity 4 times coarser, due to its poorer resolution, which limits the multipole range available for the analysis. 
The higher resolution ground based Advanced ACT experiment achieves a sensitivity on $\alpha$ which is intermediate between LiteBIRD and COrE. 
Despite its coarser angular resolution the LSPE balloon borne experiment can achieve bounds on $\alpha$ about twice as stringent as present day high resolution experiments.
The ideal full sky cases show that cosmic variance is still far from being the dominant source of error for future experiments: even for COrE noise represents the dominant source of uncertainty (again ignoring systematic effects) being at least two order of magnitude larger than cosmic variance. In fact, cosmic variance alone prevents detection of $\alpha$ to better than $0.02$ arcsec for $r=0.01$. This limit becomes even smaller assuming lower values of the tensor-to-scalar ratio.

For the case of COrE we have checked the impact of halving the accessible multipole range of the mission by setting $\ell_{max} = 1000$, finding a small impact on the error estimate of $\alpha$ (less than 10\%). In fact the highest leverage to the constraining power in $\alpha$ comes from the multipole region below $\ell = 1500$.

\section{Conclusions}\label{Conclusions}

We have applied the well known D-estimators for the birefringence angle to the recently published ACTpol data using a frequentist MC based approach. Our best estimate for the uniform birefringence angle is $\tilde{\alpha} = 0.29^\circ \pm 0.28^\circ$ (stat.) $\pm 0.5^\circ$ (syst.) compatible with no detection and in agreement with previous analyses \cite{Mei2015,Zhao2015}. The systematic error is an estimate given by the ACTpol collaboration \cite{ACTpol2014}.

We have also presented the dependence of uniform $\alpha$ on subintervals of the whole multipole range used for the global analysis. This test that was not previously known for ACTpol, provides no significant deviation from zero in any of the subintervals considered.

Furthermore, we used our MC machinery to provide forecasts for the planned experiments Advanced ACT, LSPE, LiteBIRD and as well as a possible COrE-like mission. We find that the best sensitivity is achieved by COrE that may be able to constrain the birefringence angles at a level of 10 arcsec provided that all systematic effects can be kept under control. Under the same conditions, all other experiments are expected to be at least about 2-3 times less competitive.

Even the sensitivity achievable with COrE lies a couple of order of magnitudes above what can be reasonably expected from a cosmic variance limited ideal experiment covering the full sky.

\section*{Appendix A: Signal-to-noise ratio analysis}\label{Consistency_test}

\begin{figure}
\centering
\includegraphics[scale=0.6]{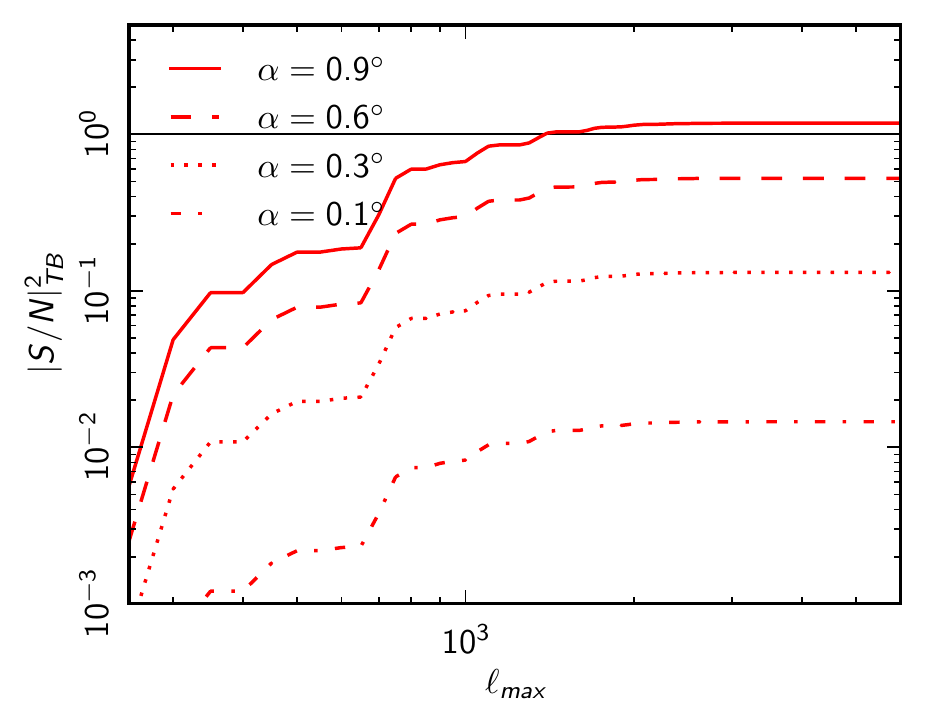}
\includegraphics[scale=0.6]{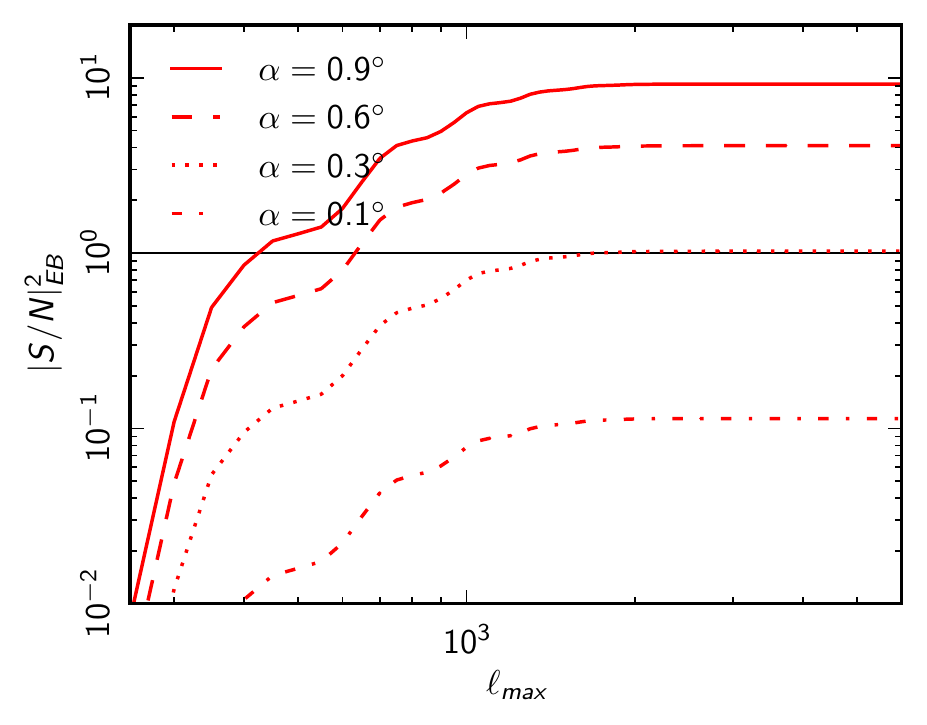}
\caption{\footnotesize{{\it Left panel}: Signal-to-noise ratio for TB as defined in Eq. (\ref{signalTB}) for ACTpol data and considering several $\alpha=0.9^\circ$ (solid), $\alpha=0.6^\circ$ (dashed), $\alpha=0.3^\circ$ (dotted), $\alpha=0.1^\circ$ (dash dot). {\it Right panel}: As in the left panel but for EB as defined in Eq. (\ref{signalEB}). }}\label{signaltonoise_ACT}
\end{figure}
\begin{figure}
\centering
\includegraphics[scale=0.6]{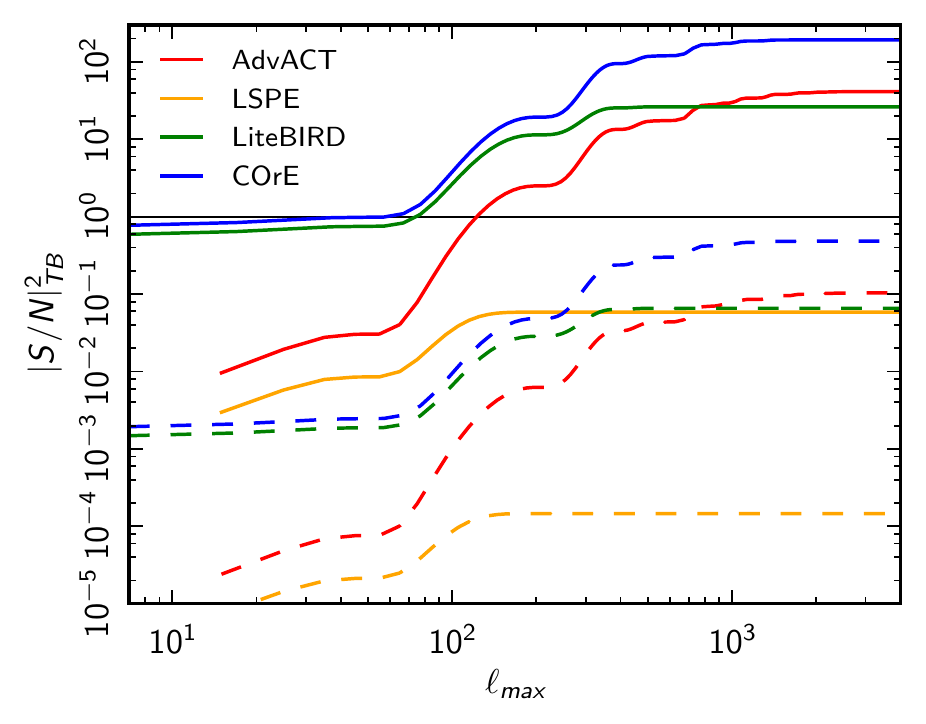}
\includegraphics[scale=0.6]{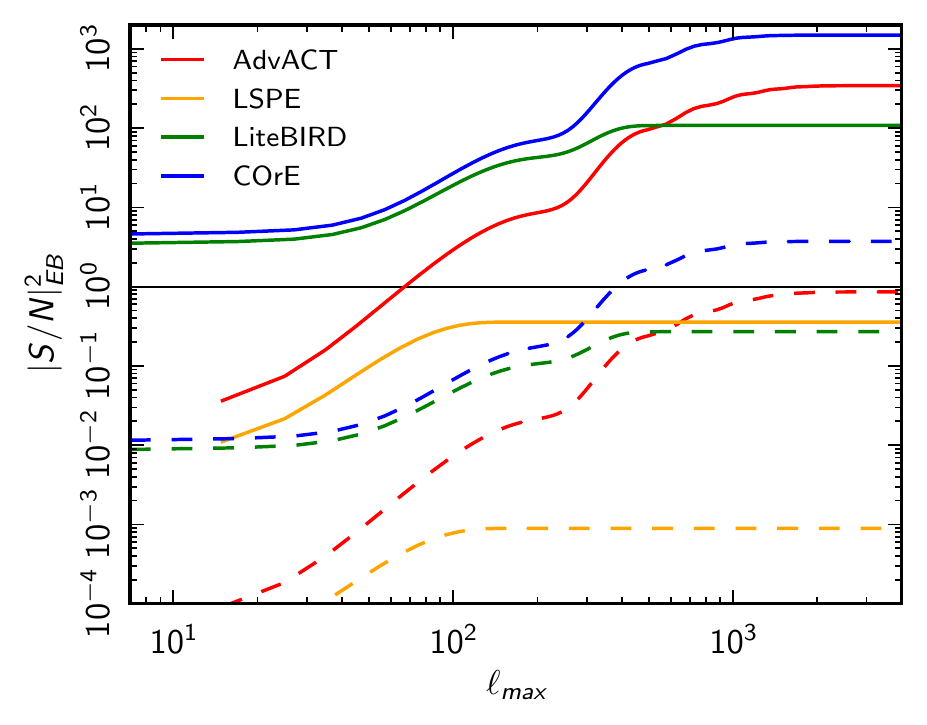}
\caption{\footnotesize{{\it Left panel}: Forecast of the signal-to-noise ratios for TB as defined in Eq. (\ref{signalTB}) for AdvACT (red lines), LSPE (orange lines), LiteBIRD (green lines) or COrE (blue lines). Two values of $\alpha$ are considered: $\alpha=0.1^\circ$ (solid lines) and $\alpha=0.005^\circ$ (dashed lines). {\it Right panel}: As in the left panel but for EB as defined in Eq. (\ref{signalEB}). }}\label{signaltonoise_Fore}
\end{figure}

In order to find the maximum multipole, $\ell_{max}$, to be considered in our analyses, we take into account the following signal-to-noise ratio functions defined as:
\begin{eqnarray}
\left(\frac{S}{N}\right)^2_{TB}(\ell_{max}) &=& \sum_{\ell=\ell_{min}}^{\ell_{max}} \left(\frac{C_{\ell}^{TE} sin(2\alpha)}{\sigma_{\ell}^{TB}}\right)^2\label{signalTB} \\
\left(\frac{S}{N}\right)^2_{EB}(\ell_{max}) &=& \sum_{\ell=\ell_{min}}^{\ell_{max}} \left(\frac{1}{2} \frac{(C_{\ell}^{EE}-C_{\ell}^{BB})sin(4\alpha)}{\sigma_{\ell}^{EB}}\right)^2
\label{signalEB}
\end{eqnarray}
where $C_{\ell}^{TE}$, $C_{\ell}^{EE}$ and $C_{\ell}^{BB}$ are chosen to be the best fit model obtained from Planck 2015 \cite{planck2014-a15} and where $\sigma_{TB}$ and $\sigma_{EB}$ are the uncertainties for the TB and EB spectra respectively. 

Fig. \ref{signaltonoise_ACT} shows Eq. (\ref{signalTB}) (left panel) and (\ref{signalEB}) (right panel) for ACTpol data and considering $\alpha=0.1^\circ$, $0.3^\circ$, $0.6^\circ$ and $0.9^\circ$.
Both left and right panel of Fig. \ref{signaltonoise_ACT} exhibit a saturation of the signal-to-noise ratio at $\ell_{max}$ of the order of $3000$.
This justifies the choice of $\ell_{max}=3000$ when we perform our analyses of ACTpol data, see Section \ref{Results}.

Similarly Fig. \ref{signaltonoise_Fore} shows Eq. (\ref{signalTB}) and (\ref{signalEB}) for future CMB experiments as listed in Table \ref{table} considering $\alpha = 0.1^\circ$ and $0.005^\circ$. It clearly displays that the saturation happens at $\ell_{max} \sim 3000$ for AdvACT, $\ell_{max} \sim 200$ for LSPE, $\ell_{max} \sim 600$ for LiteBIRD and $\ell_{max} \sim 2000$ for COrE. Therefore we choose the maximum multipole accordingly in our forecasts, see Section \ref{Forecasts}.

\section*{Acknowledgments}

We acknowledge the support by ASI/INAF Agreement 2014-024-R.1 for the Planck LFI Activity of Phase E2.

\bibliography{mybibfile}

\begin{thebibliography}{10}
\expandafter\ifx\csname url\endcsname\relax
  \def\url#1{\texttt{#1}}\fi
\expandafter\ifx\csname urlprefix\endcsname\relax\def\urlprefix{URL }\fi
\expandafter\ifx\csname href\endcsname\relax
  \def\href#1#2{#2} \def\path#1{#1}\fi

\bibitem{planck2014-a01}
{Planck Collaboration}, {\textit{Planck} 2015 results. I. Overview of products
  and results}, A\&A 594 (2016) A1.
\newblock \href {http://arxiv.org/abs/1502.01582} {\path{arXiv:1502.01582}}.

\bibitem{pb2015}
{BICEP2/Keck Array and Planck Collaborations}, {Joint Analysis of BICEP2/Keck
  Array and Planck Data}, Physical Review Letters 114~(10) (2015) 101301.

\bibitem{ACT2013}
J.~L. {Sievers}, R.~A. {Hlozek}, M.~R. {Nolta}, {et al.}, {The Atacama
  Cosmology Telescope: cosmological parameters from three seasons of data},
  JCAP 10 (2013) 60.

\bibitem{SPT2011}
K.~K. {Schaffer}, T.~M. {Crawford}, K.~A. {Aird}, {et al.}, {The First Public
  Release of South Pole Telescope Data: Maps of a 95 deg$^{2}$ Field from 2008
  Observations}, APJ 743 (2011) 90.

\bibitem{planck2014-a13}
{Planck Collaboration}, {\textit{Planck} 2015 results. XI. CMB power spectra,
  likelihoods, and robustness of parameters}, A\&A 594 (2016) A11.
\newblock \href {http://arxiv.org/abs/1507.02704} {\path{arXiv:1507.02704}}.

\bibitem{AdvACTpol2015}
S.~W. {Henderson}, R.~{Allison}, J.~{Austermann}, {et al.}, {Advanced ACTPol
  Cryogenic Detector Arrays and Readout~}, Journal of Low Temperature Physics
  184 (2016) 772--779.
\newblock \href {http://arxiv.org/abs/1510.02809} {\path{arXiv:1510.02809}}.

\bibitem{SPTpol2014}
E.~M. {George}, C.~L. {Reichardt}, K.~A. {Aird}, {et al.}, {A Measurement of
  Secondary Cosmic Microwave Background Anisotropies from the 2500
  Square-degree SPT-SZ Survey}, APJ 799 (2015) 177.

\bibitem{Bicep32014}
Z.~{Ahmed}, M.~{Amiri}, S.~J. {Benton}, {et al.}., {BICEP3: a 95GHz refracting
  telescope for degree-scale CMB polarization}, in: Society of Photo-Optical
  Instrumentation Engineers (SPIE) Conference Series, Vol. 9153 of Society of
  Photo-Optical Instrumentation Engineers (SPIE) Conference Series, 2014, p.~1.
\newblock \href {http://arxiv.org/abs/1407.5928} {\path{arXiv:1407.5928}}.

\bibitem{ACTpol2014}
S.~{Naess}, M.~{Hasselfield}, J.~{McMahon}, {et al.}, {The Atacama Cosmology
  Telescope: CMB polarization at 200 $\leq$ l $\leq$ 9000}, JCAP 10 (2014) 7.

\bibitem{SPTpol2015}
K.~T. {Story}, D.~{Hanson}, P.~A.~R. {Ade}, {et al.}, {A Measurement of the
  Cosmic Microwave Background Gravitational Lensing Potential from 100 Square
  Degrees of SPTpol Data}, APJ 810 (2015) 50.

\bibitem{LSPE2012}
{The LSPE collaboration}, {The Large-Scale Polarization Explorer (LSPE),~}\href
  {http://arxiv.org/abs/1208.0281} {\path{arXiv:1208.0281}}.

\bibitem{Litebird2014}
T.~{Matsumura}, Y.~{Akiba}, J.~{Borrill}, {et al.}, {Mission Design of
  LiteBIRD}, Journal of Low Temperature Physics 176 (2014) 733--740.

\bibitem{Core+2014}
J.~{Delabrouille}, {COrE+ The Cosmic Origins Explorer A proposal for ESA's M4
  space mission}, Retrieved from the University of Minnesota Digital
  Conservancy, http://hdl.handle.net/11299/169642.

\bibitem{Carroll1990}
S.~M. {Carroll}, G.~B. {Field}, R.~{Jackiw}, {Limits on a Lorentz- and
  parity-violating modification of electrodynamics}, Physical Review D 41
  (1990) 1231--1240.

\bibitem{Carroll1998}
S.~M. {Carroll}, {Quintessence and the Rest of the World: Suppressing
  Long-Range Interactions}, Physical Review Letters 81 (1998) 3067--3070.

\bibitem{Giovannini2005}
M.~{Giovannini}, {Magnetized birefringence and CMB polarization}, Physical
  Review D 71~(2) (2005) 021301.

\bibitem{Finelli2009}
F.~{Finelli}, M.~{Galaverni}, {Rotation of linear polarization plane and
  circular polarization from cosmological pseudoscalar fields}, Physical Review
  D 79~(6) (2009) 063002.

\bibitem{Bartolo2015}
N.~{Bartolo}, S.~{Matarrese}, M.~{Peloso}, M.~{Shiraishi}, {Parity-violating
  CMB correlators with non-decaying statistical anisotropy}, JCAP 7 (2015) 39.

\bibitem{PVLAS2016}
F.~{Della Valle}, A.~{Ejlli}, U.~{Gastaldi}, G.~{Messineo}, E.~{Milotti},
  R.~{Pengo}, G.~{Ruoso}, G.~{Zavattini}, {The PVLAS experiment: measuring
  vacuum magnetic birefringence and dichroism with a birefringent Fabry-Perot
  cavity}, European Physical Journal C 76 (2016) 24.

\bibitem{Alighieri2010}
S.~{di Serego Alighieri}, F.~{Finelli}, M.~{Galaverni}, {Cosmological
  birefringence: an astrophysical test of fundamental physics}, in: PPC 2010:
  IV International Workshop On The Interconnection Between Particle Physics And
  Cosmology, Proceedings of the conference held 12-16 July, 2010 in Torino,
  Italy. Online at http://www.ppc10.to.infn.it/talks.php, id.27, 2010, p.~27.
\newblock \href {http://arxiv.org/abs/1011.4865} {\path{arXiv:1011.4865}}.

\bibitem{Kamionkowski2010}
M.~{Kamionkowski}, {Nonuniform cosmological birefringence and active galactic
  nuclei}, Physical Review D 82~(4) (2010) 047302.

\bibitem{Lue1999}
A.~{Lue}, L.~{Wang}, M.~{Kamionkowski}, {Cosmological Signature of New
  Parity-Violating Interactions}, Physical Review Letters 83 (1999) 1506--1509.

\bibitem{Li2009}
M.~{Li}, Y.-F. {Cai}, X.~{Wang}, X.~{Zhang}, {CPT violating electrodynamics and
  Chern-Simons modified gravity}, Physics Letters B 680 (2009) 118--124.

\bibitem{Alghieri2015}
S.~{di Serego Alighieri}, {Cosmic Polarization Rotation in view of the recent
  CMB experiments}, ArXiv e-prints~\href {http://arxiv.org/abs/1507.02433}
  {\path{arXiv:1507.02433}}.

\bibitem{Kaufman2016}
J.~P. {Kaufman}, B.~G. {Keating}, B.~R. {Johnson}, {Precision tests of parity
  violation over cosmological distances}, MNRAS 455 (2016) 1981--1988.

\bibitem{Contaldi2015}
C.~R. {Contaldi}, {Imaging parity-violation in the CMB,~}\href
  {http://arxiv.org/abs/1510.02629} {\path{arXiv:1510.02629}}.

\bibitem{Gruppuso2015}
A.~{Gruppuso}, M.~{Gerbino}, P.~{Natoli}, L.~{Pagano}, N.~{Mandolesi},
  D.~{Molinari}, {Constraints on cosmological birefringence from Planck and
  Bicep2/Keck data}, JCAP 6 (2016) 1.
\newblock \href {http://arxiv.org/abs/1509.04157} {\path{arXiv:1509.04157}}.

\bibitem{Gluscevic2012}
V.~{Gluscevic}, D.~{Hanson}, M.~{Kamionkowski}, C.~M. {Hirata}, {First CMB
  constraints on direction-dependent cosmological birefringence from WMAP-7},
  Phys. Rev. D 86~(10).

\bibitem{Polarbear2015}
{Polarbear Collaboration}, P.~A.~R. {Ade}, K.~{Arnold}, M.~{Atlas},
  C.~{Baccigalupi}, {et al.}, {POLARBEAR constraints on cosmic birefringence
  and primordial magnetic fields}, Phys. Rev. D 92~(12).

\bibitem{Zhaobis2014}
W.~{Zhao}, M.~{Li}, {Fluctuations of cosmological birefringence and the effect
  on CMB B-mode polarization}, Physical Review D 89~(10) (2014) 103518.

\bibitem{Zhao2014}
W.~{Zhao}, M.~{Li}, {Detecting relic gravitational waves in the CMB: The
  contamination caused by the cosmological birefringence}, Physics Letters B
  737 (2014) 329--334.

\bibitem{Gluscevic2009}
V.~{Gluscevic}, M.~{Kamionkowski}, A.~{Cooray}, {Derotation of the cosmic
  microwave background polarization: Full-sky formalism}, Phys. Rev. D 80~(2).

\bibitem{Yadav2009}
A.~P.~S. {Yadav}, R.~{Biswas}, M.~{Su}, M.~{Zaldarriaga}, {Constraining a
  spatially dependent rotation of the cosmic microwave background
  polarization}, Phys. Rev. D 79~(12).

\bibitem{Wu2009}
E.~Y.~S. {Wu}, P.~{Ade}, J.~{Bock}, {et al.}, {Parity Violation Constraints
  Using Cosmic Microwave Background Polarization Spectra from 2006 and 2007
  Observations by the QUaD Polarimeter}, Physical Review Letters 102~(16)
  (2009) 161302.

\bibitem{Gruppuso2012}
A.~{Gruppuso}, P.~{Natoli}, N.~{Mandolesi}, A.~{De Rosa}, F.~{Finelli},
  F.~{Paci}, {WMAP 7 year constraints on CPT violation from large angle CMB
  anisotropies}, JCAP 2 (2012) 023.

\bibitem{Zhao2015}
G.-B. {Zhao}, Y.~{Wang}, J.-Q. {Xia}, M.~{Li}, X.~{Zhang}, {An efficient probe
  of the cosmological CPT violation}, JCAP 7 (2015) 32.

\bibitem{Gruppuso2016}
A.~{Gruppuso}, G.~{Maggio}, D.~{Molinari}, P.~{Natoli}, {A note on the
  birefringence angle estimation in CMB data analysis}, JCAP 5 (2016) 20.
\newblock \href {http://arxiv.org/abs/1604.05202} {\path{arXiv:1604.05202}}.

\bibitem{PrivateComm}
S.~{Naess}, Private communication.

\bibitem{planck2014-a15}
{Planck Collaboration}, {\textit{Planck} 2015 results. XIII. Cosmological
  parameters}, A\&A 594.
\newblock \href {http://arxiv.org/abs/1502.01589} {\path{arXiv:1502.01589}}.

\bibitem{Zaldarriaga1997}
M.~{Zaldarriaga}, U.~{Seljak}, {All-sky analysis of polarization in the
  microwave background}, Physical Review D 55 (1997) 1830--1840.

\bibitem{Kamionkowski1997}
M.~{Kamionkowski}, A.~{Kosowsky}, A.~{Stebbins}, {Statistics of cosmic
  microwave background polarization}, Physical Review D 55 (1997) 7368--7388.

\bibitem{Mei2015}
H.-H. {Mei}, W.-T. {Ni}, W.-P. {Pan}, L.~{Xu}, S.~{di Serego Alighieri}, {New
  Constraints on Cosmic Polarization Rotation from the ACTPol Cosmic Microwave
  Background B-mode Polarization Observation and the BICEP2 Constraint Update},
  APJ 805 (2015) 107.

\end{thebibliography}

\end{document}